\documentclass[english,11pt,a4paper]{article}   

\usepackage[english]{babel}		
\usepackage[utf8]{inputenc}
\usepackage[T1]{fontenc}	
\usepackage{multicol}
\usepackage{graphicx}
\usepackage{mathtools}	
\usepackage{siunitx}
\usepackage{xcolor}
\usepackage{url}
\usepackage{hyperref}
\usepackage{amssymb}
\usepackage{varioref}
\usepackage[draft]{fixme}
\usepackage{lipsum}
\usepackage{nth}
\usepackage{appendix}
\usepackage{cite}
\usepackage{amsmath}
\usepackage{amsthm}
\usepackage{fixme}
\usepackage{soul}
\usepackage{float}
\usepackage{placeins}
\usepackage{wrapfig}
\usepackage{scalefnt}
\usepackage{framed}
\usepackage[a4paper, total={6in, 8in}]{geometry}
\usepackage{authblk}
\usepackage{tikz}
\usepackage{pgf,tikz}
\usetikzlibrary{decorations.pathreplacing}
\usepackage{pgfplots}
\usetikzlibrary{arrows, positioning, calc}

\usepackage{chngcntr}

\newcounter{fpcounter}

\vrefwarning

\title{Surveying the Attitudes of Physicists Concerning Foundational Issues of Quantum Mechanics}

\author[1]{Sujeevan Sivasundaram \thanks{sujeevan.sivasundaram@post.au.dk}}
\author[2]{Kristian Hvidtfelt Nielsen}
\affil[1]{Department of Physics and Astronomy, Aarhus University}
\affil[2]{Center for Science Studies, Aarhus University}

\date{}

\begin{document}

\maketitle

\begin{abstract}
Even though quantum mechanics has existed for almost 100 years, questions concerning the foundation and interpretation of the theory still remain. These issues have gathered more attention in recent years, but does this mean that physicists are more aware of foundational issues concerning quantum mechanics? A survey was sent out to 1234 physicists affiliated to 8 different universities. 149 responded to the questions, which both concerned foundational issues related to quantum mechanics, specifically, as well as questions concerning interpretations of physical theories in general. The answers to the survey revealed that foundational concepts in quantum mechanics are still a topic that only a minority of physicists are familiar with, although a clear majority of physicists find that interpretations of physical theories are important. The various questions, as well as how the respondents answered, are presented. The survey intends to give an overview of what the opinion of the physics community, in general, is concerning issues related to quantum mechanics.

\end{abstract}



\section{Introduction}
Though quantum mechanics is arguably the most successful theory in physics, its formalism does not lend itself easily to an interpretation, that would make it possible to envision the various processes described by the theory. This issue has plagued quantum mechanics since its conception, but the issue is very distinct from other issues which can plague physical theories. Some would even claim that the issue has been solved long ago, while others would refute the whole notion of it being an issue at all. This issue, or this non-issue if you like, confronts physics with the question of what is required of a physical theory? Is it enough that its formalism is able to make correct predictions or does it need to give an explicable description of what is being described by the formalism? These two options represent different ends of a spectrum but are at the same time not completely distinct. A theory that gives correct predictions must surely describe some aspect of nature in some way, though it may not be mediated through human language or through pictures we can imagine. 


Once a subject that would doom a physicist's career, should he engage in it, quantum foundations seem to have gained popularity as a research subject, and today there are a plethora of interpretations of quantum mechanics. The interpretations in this context are in reality different theories that are designed to replicate the same results as standard quantum mechanics but solve some foundational issues such as the measurement problem for example. These different interpretations cannot be separated by experiments\footnote{There are exceptions to this statement, such as the GRW-collapse theory, that can be refuted through experiments, which are presently not possible, but could be in the future.}, since they are designed to give the same predictions. How should physicists then choose between the different interpretations? And is this a question that physics should concern itself with?

The survey was carried out in relation to a master thesis project carried out at Aarhus University. The full thesis can be found here: \href{http://css.au.dk/fileadmin/reposs/reposs-039.pdf}{http://css.au.dk/fileadmin/reposs/reposs-039.pdf}. 

\section{Survey: Uncovering the Attitudes of Physicists}
The answers to such questions are not easily found, and may not even exist as pure answers, where one can distinguish right from wrong, but may only exist in the form of opinions. However, these opinions may shape how physics will and can move forward in the future. To uncover the current landscape of opinions and attitudes to these questions a survey was carried out, much inspired from that of Schlosshauer et al. in 2013 \cite{Article}. Many of the same questions were used, some were altered and some new questions were added. Unlike Schlosshauer et Al. the survey was not exclusively given to experts in quantum foundations but was given to all kinds of physicists from Aarhus University, Copenhagen University, Göttingen University, Heidelberg University, Oxford University, California Institute of Technology, National University Singapore and University College London. The choice of the universities was mostly arbitrary, though certain factors did influence the choice. One factor was their connection to Aarhus University since it was thought that more people would participate in the survey if it came from a university they knew well. Another factor was whether the university had a relation to the development of quantum mechanics, which would perhaps make people more inclined to answer from a sense of heritage. A last factor was simply a case of logistics; how easy or difficult it was to harvest the email addresses from the various universities’ websites.

A personal link to the questionnaire was sent by email to 1234 physicists, who were either graduate students, Ph.D. students, Ph..D graduates, Professors or Lector Emeriti. Out of 1234, only 150 participated in the survey, corresponding to about 12\% answering the survey. One of these participants did not answer the online survey but wrote an email with his opinions and answers\footnote{The participant chose other options than those given, it was therefore not possible to incorporate his answers.}, so there are results from 149 of the participants. 

Other surveys of this nature, besides Scholsshauer et Al., have been carried out with varied results \cite{Snap1} \cite{Snap2} \cite{Snap3}. The survey here has significantly more participants than any of those referred to, and since it is distributed worldwide, it should give a more representative view of the opinions of physicists. The participants consist largely of Danes with about 44\% of participants having Danish nationality.

\section{The Questions and What Was Answered}

Questions 1,2,5 and 17 were taken from Schlosshauer et al. and used unaltered in the questionnaire, while questions 4,6,7 and were also taken from Scholsshauer et al., but were altered slightly either in their formulation or the answer-options. 

\vspace{5mm}

\begin{figure}[H]
\includegraphics[scale=0.8]{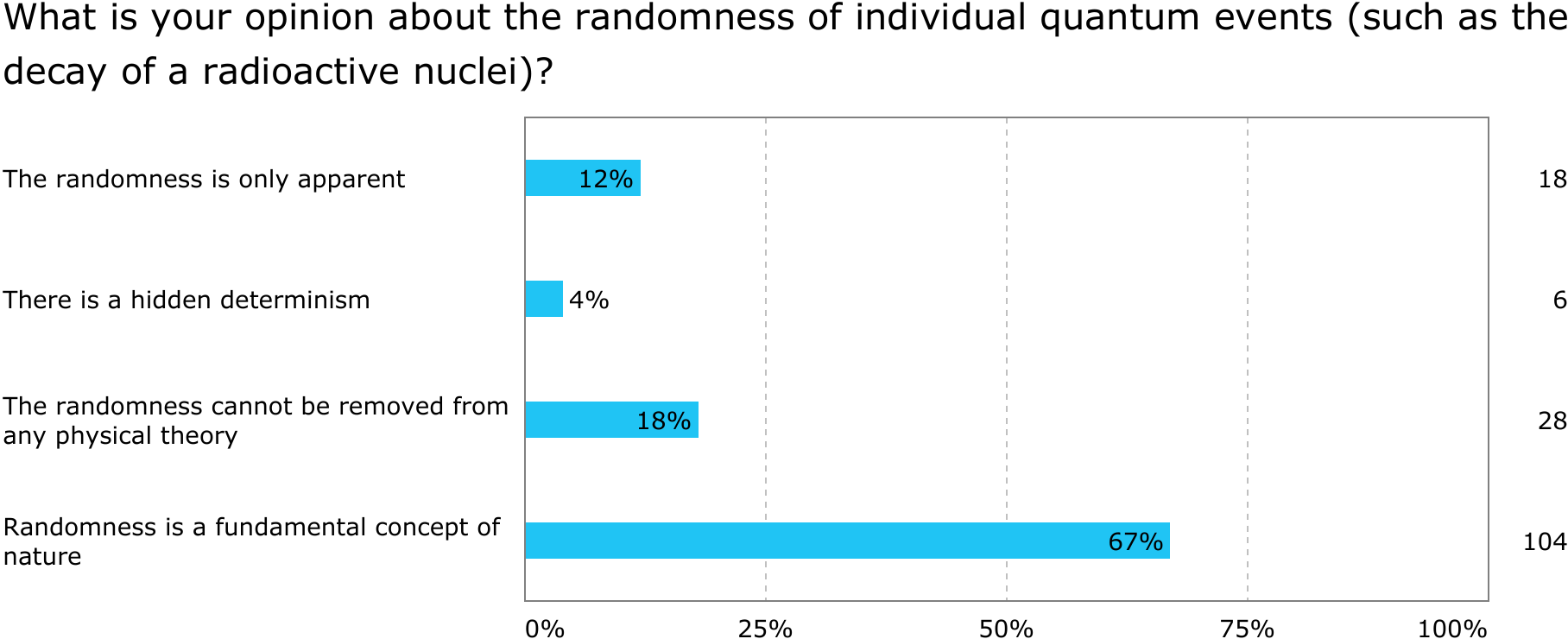}
\caption{Distribution of all the participants answers to question 1}
\label{Q1}
\end{figure}

The first question is intended to investigate the specific opinions regarding the randomness found in quantum mechanics. The various answers correspond to how one would answer the question from the viewpoint of different interpretations. Thus, the first option "\textit{The randomness is only apparent}" corresponds to the answer one would give from the viewpoint of the many worlds interpretation, since the universal wave function evolves in a deterministic (non-random) way through the wave equation, but every observer is embedded in the universe moving along different branches giving rise to an apparent randomness from the observer's point of view. 
The second option corresponds to the answer one would give from the viewpoint of bohmian mechanics, where the observed randomness of quantum systems is only due to a lack of knowledge of the exact initial conditions. 

\begin{figure}[H]
\includegraphics[scale=0.8]{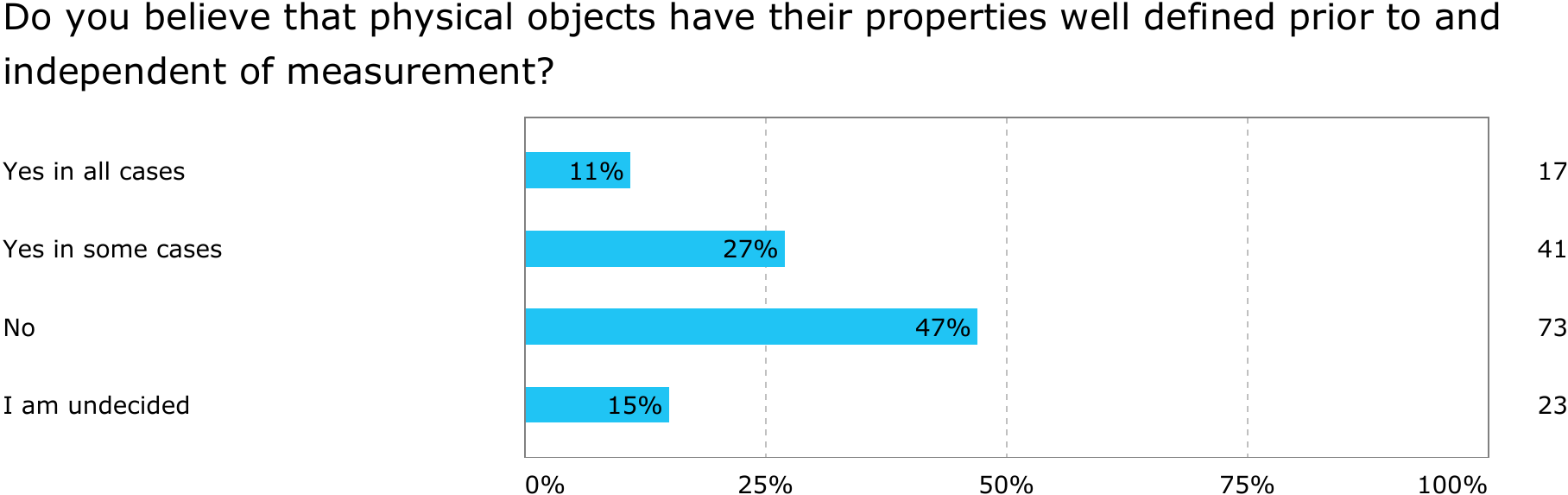}
\caption{Distribution of all the participants answers to question 2}
\label{Q2}
\end{figure}

The second question pertains to the role of measurement in defining physical properties. This question has some ambiguity to it because it might not be well-defined, what is meant by the word "\textit{physical property}".

The intention of the question was to ascertain the participants’ view of wave function collapse; is it a description of nature or our knowledge of a system? A more formal version of "Is the moon there when you are not looking?"\footnote{Allegedly Einstein posed this question in objection to the notion of collapse upon observation \cite{moon}}.

\begin{figure}[H]
\includegraphics[scale=0.8]{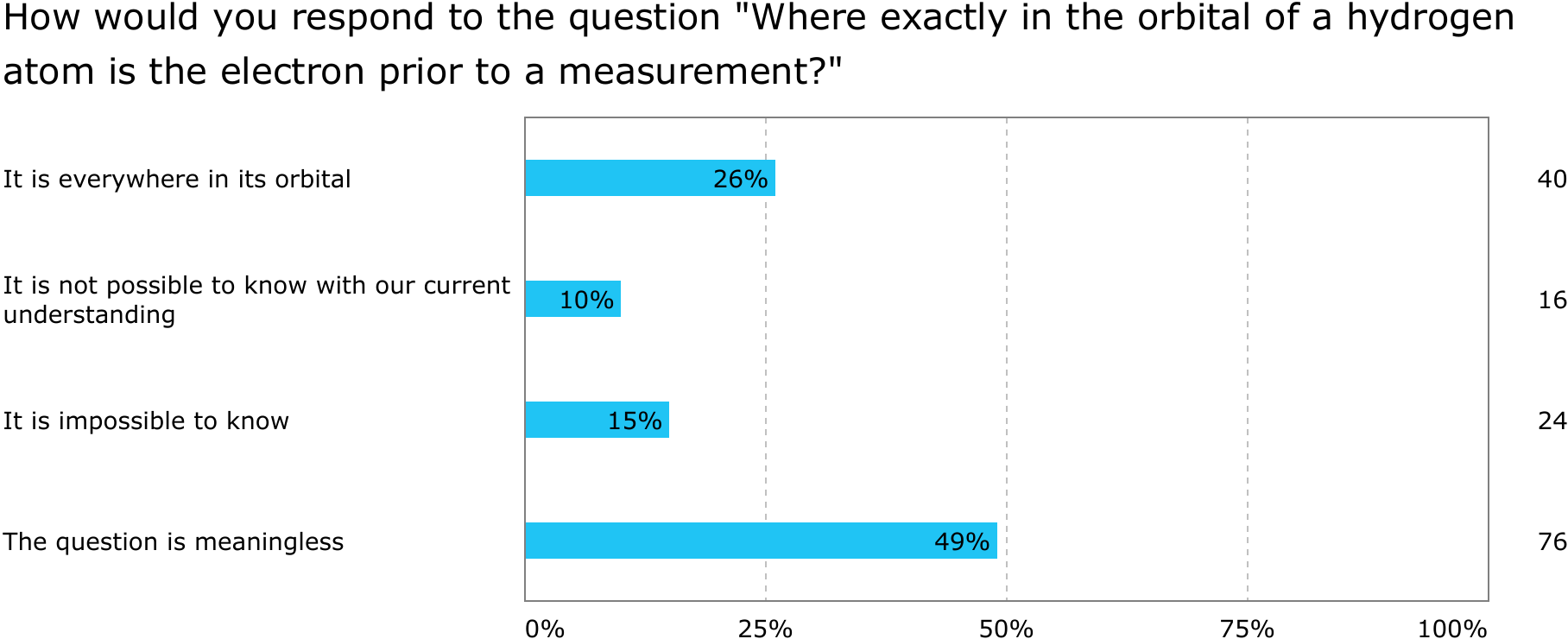}
\caption{Distribution of all the participants answers to question 3}
\label{Q3}
\end{figure}

The third question can be seen as a specific case of question 2, with the physical property being the position.

\begin{figure}[H]
\includegraphics[scale=0.8]{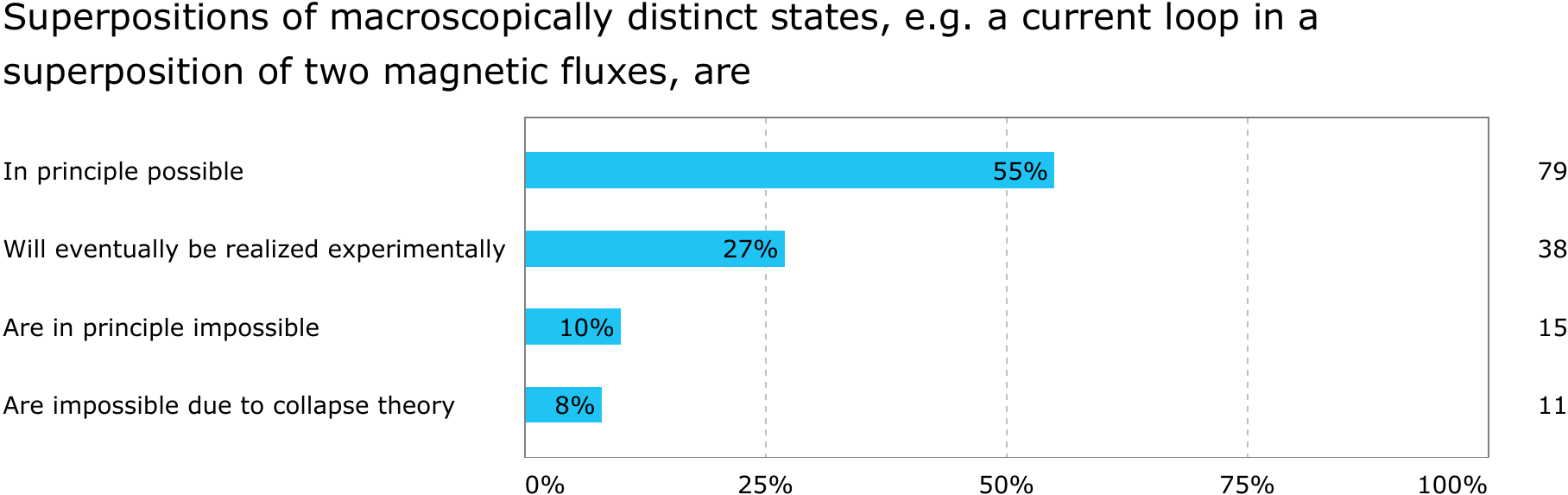}
\caption{Distribution of all the participants answers to question 4}
\label{Q4}
\end{figure}

The fourth question concerns quantum effects in macroscopic objects.

\begin{figure}[H]
\includegraphics[scale=0.8]{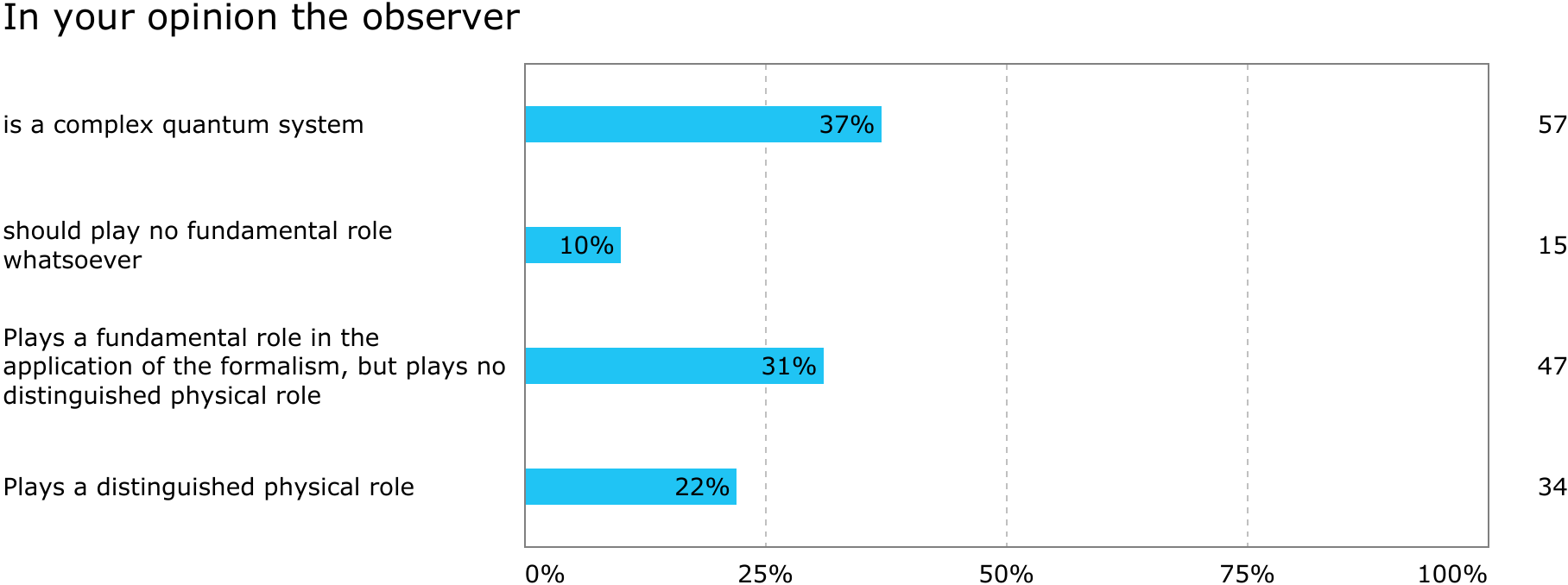}
\caption{Distribution of all the participants answers to question 5}
\label{Q5}
\end{figure}

The fifth question concerns the role the observer plays in nature.

\begin{figure}[H]
\includegraphics[scale=0.8]{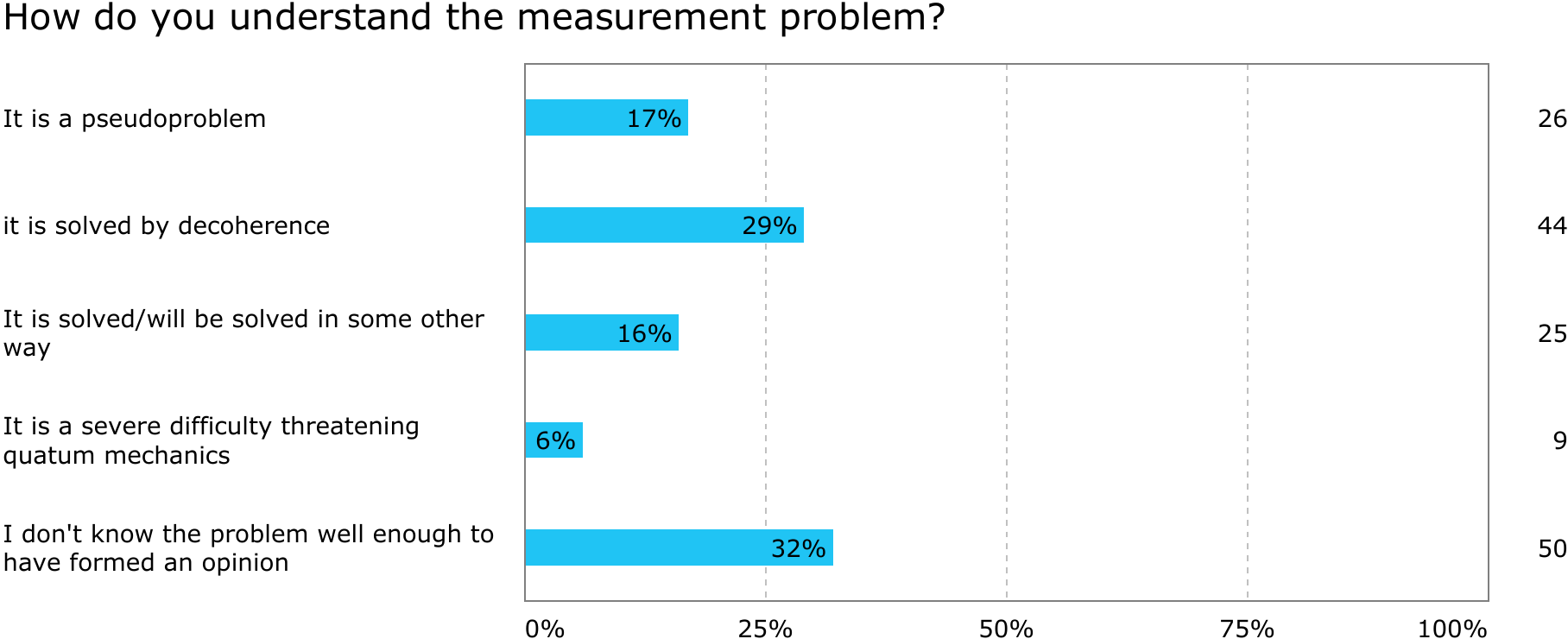}
\caption{Distribution of all the participants answers to question 6}
\label{Q6}
\end{figure}

The sixth question concerns the measurement problem. The measurement problem is often portrayed as "the" problem of the Copenhagen interpretations. The results here a very striking; the majority of the participants are not familiar with the measurement problem. This gives an indication of what role foundations of quantum mechanics play in the mind of physicists; not a significant one.

\begin{figure}[H]
\includegraphics[scale=0.8]{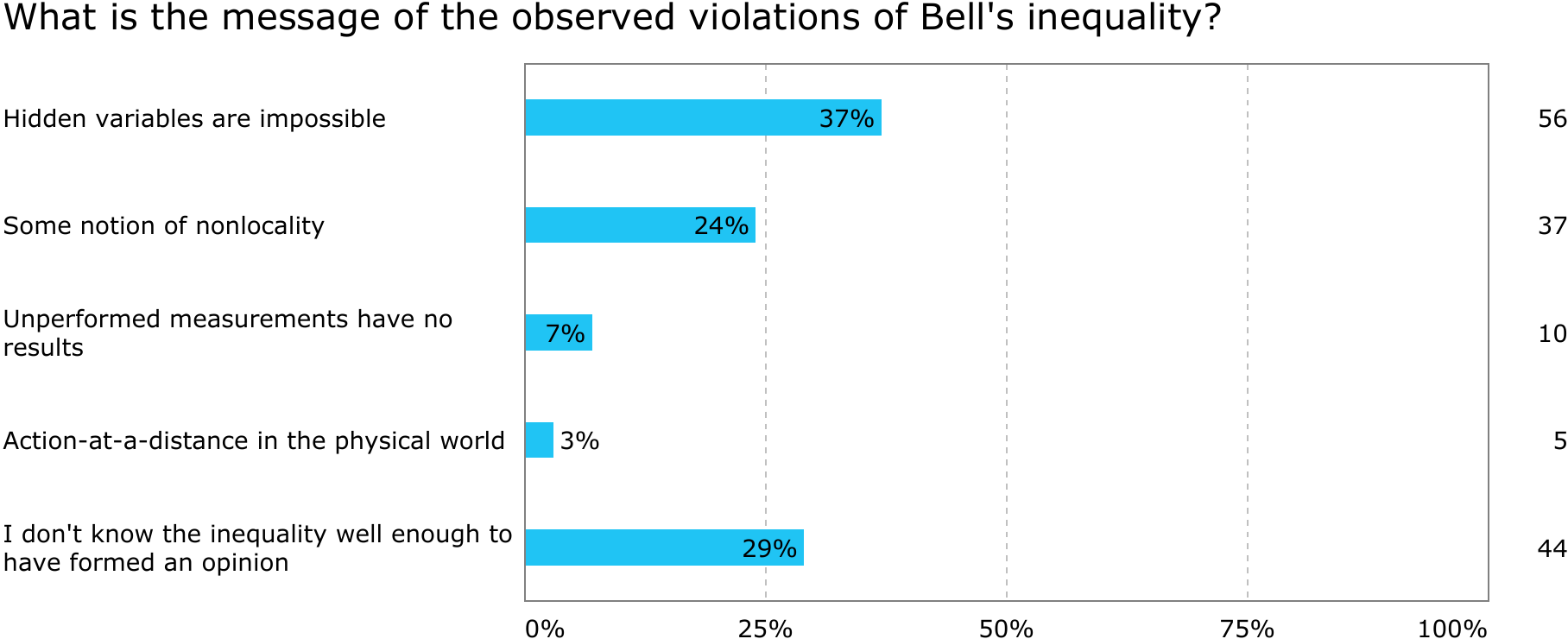}
\caption{Distribution of all the participants answers to question 7}
\label{Q7}
\end{figure}

The seventh question concerns Bell's inequality and its implication. 
Once again the results show that a significant part of the participants does not know of a concept and result pertaining to the foundations of quantum mechanics. The majority understands the violations of Bell's inequality as excluding the possibility of hidden variables, which is not true, it excludes the possibility of local hidden variables. Furthermore, 29\% of the participants do not know the inequality, which means that two-thirds of the participants do not have a proper knowledge of Bell's inequality.

\begin{figure}[H]
\includegraphics[scale=0.8]{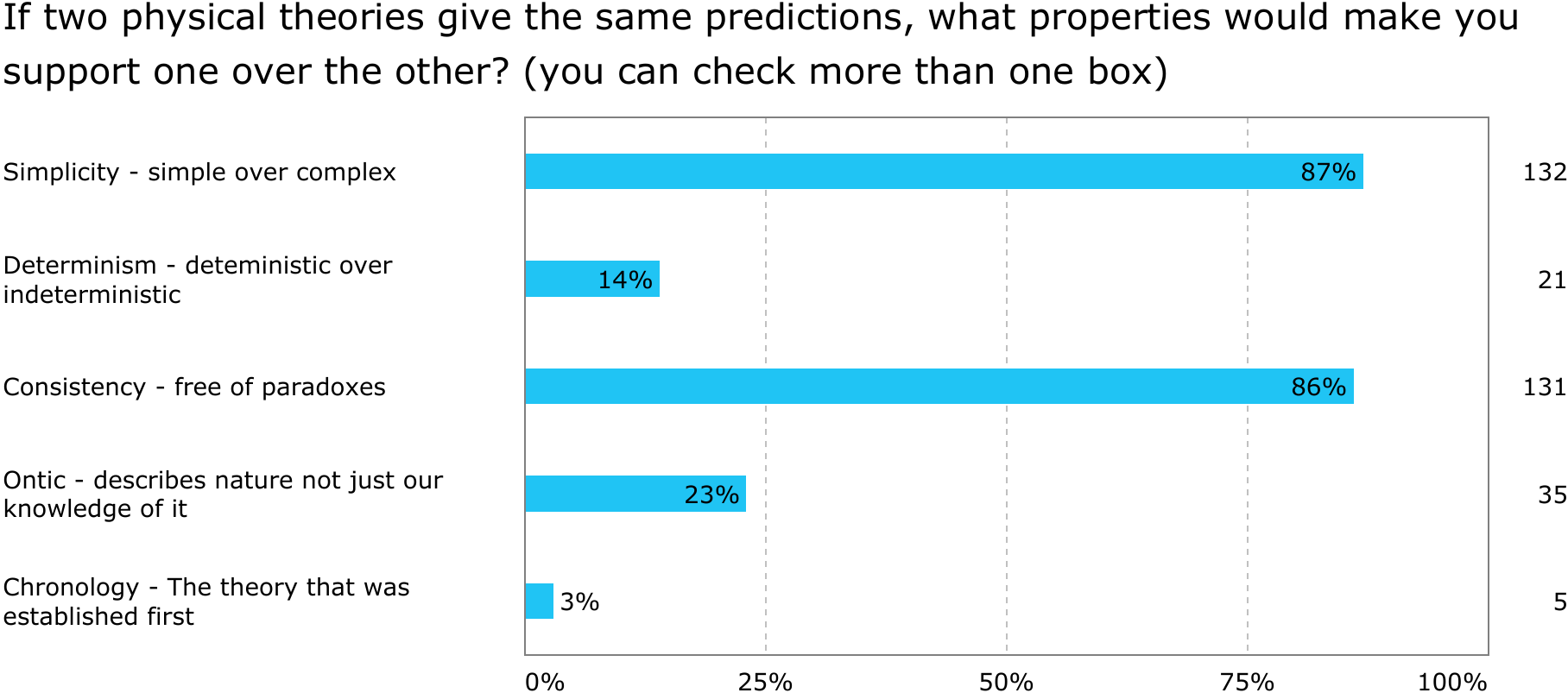}
\caption{Distribution of all the participants answers to question 8}
\label{Q8}
\end{figure}

The eighth question concerns what makes a good physical theory, specifically what makes one theory superior to another? This question does not only pertain to quantum mechanics, but to physics in general. It was allowed that the participants could pick several options in this question. The answers of the participants showed that a clear majority values the properties; simplicity and consistency. It is worth noting that so few have chosen properties as determinism and especially ontology. The answers show a divergence from the properties of classical theories.

\begin{figure}[H]
\includegraphics[scale=0.8]{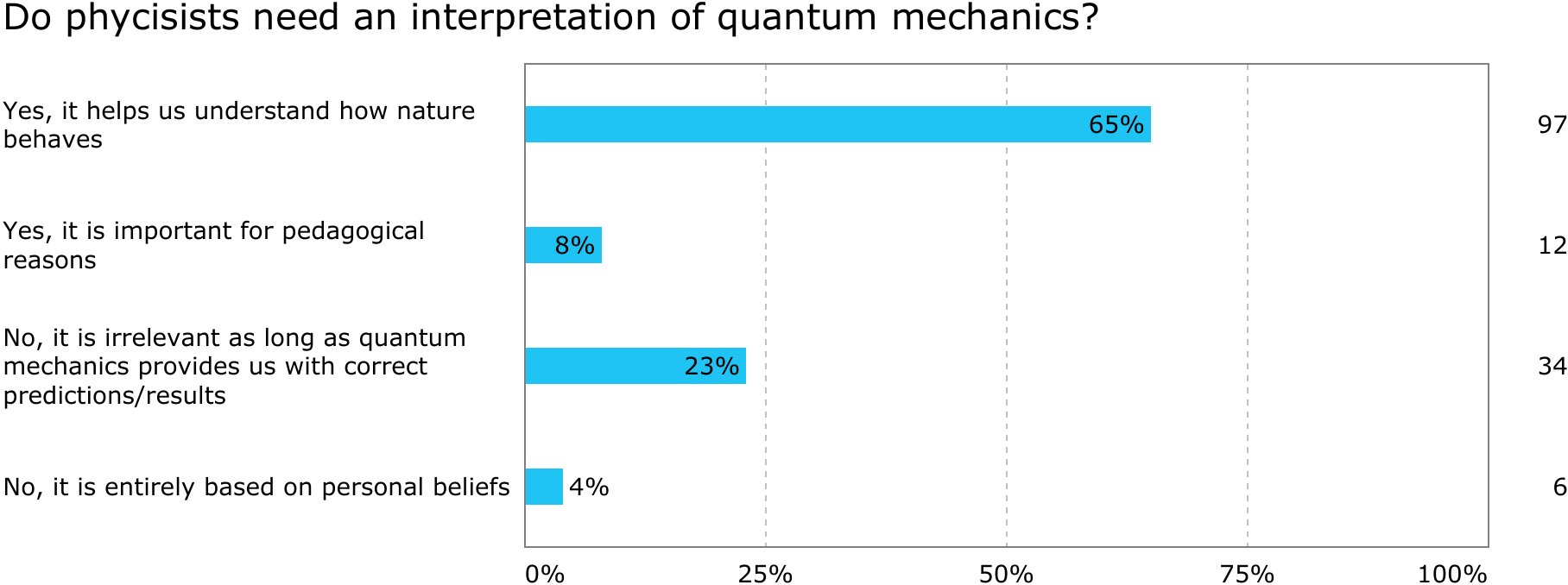}
\caption{Distribution of all the participants answers to question 9}
\label{Q9}
\end{figure}

The ninth question concerns the role a physical interpretation plays and whether it is something physicists need. There is a clear majority who feel that interpretations are necessary since it helps us describe nature. This seems quite at odds with the fact that only a fourth value an ontological theory.

\begin{figure}[H]
\includegraphics[scale=0.8]{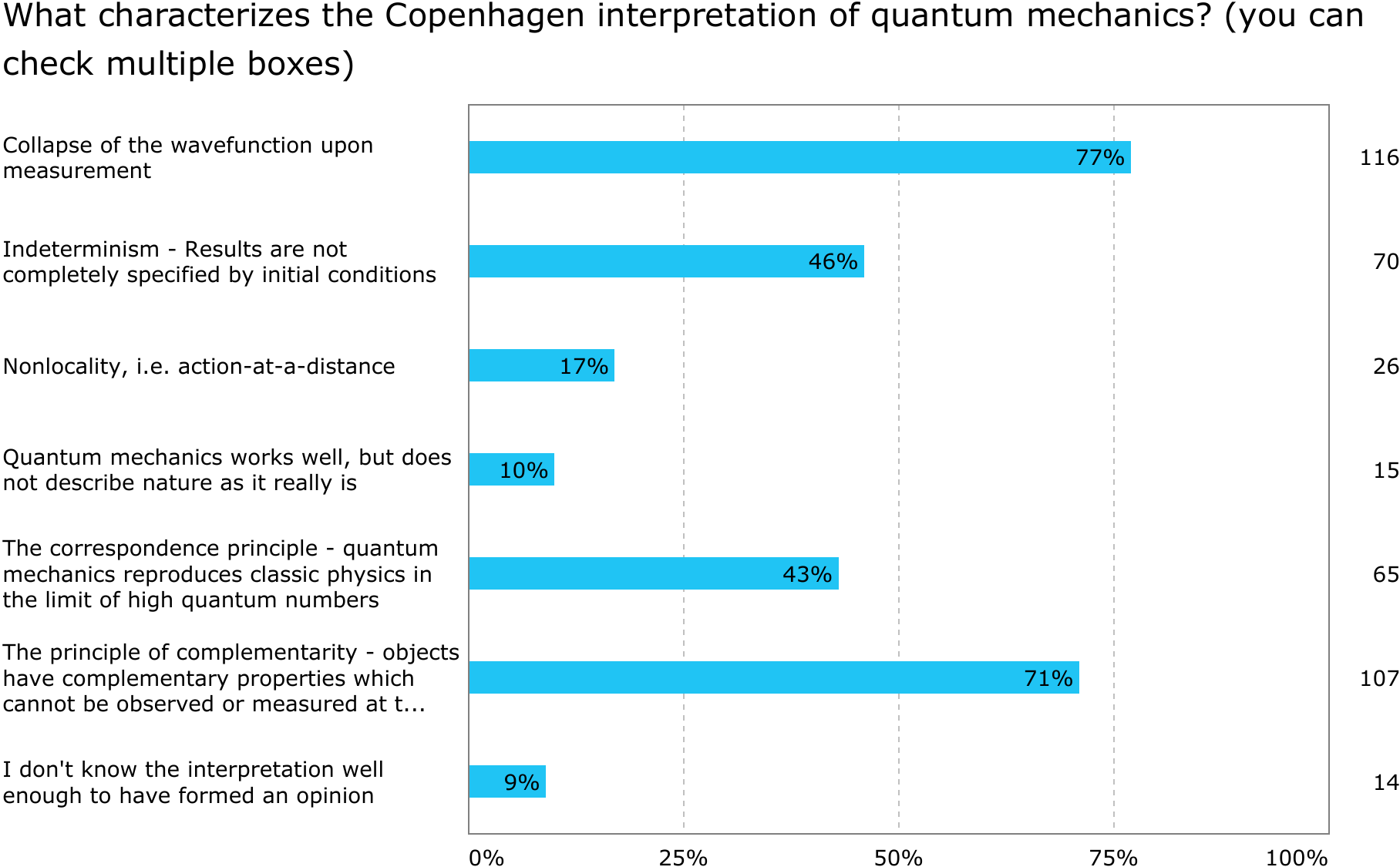}
\caption{Distribution of all the participants answers to question 10}
\label{Q10}
\end{figure}

The tenth question concerns the Copenhagen interpretation and intends to uncover what physicist associate with the Copenhagen interpretation. The participants were allowed to pick multiple options in this question.

\begin{figure}[H]
\includegraphics[scale=0.8]{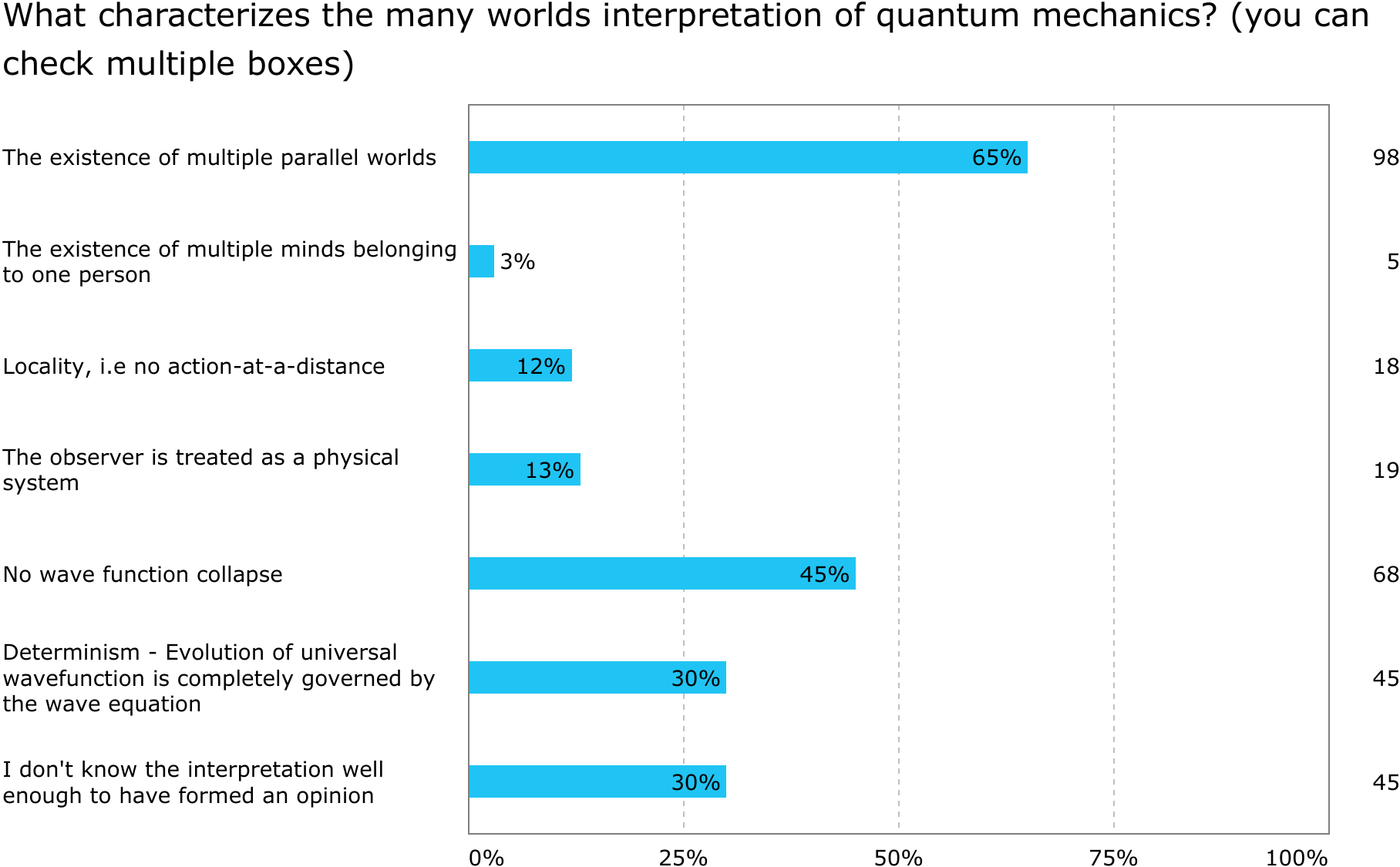}
\caption{Distribution of all the participants answers to question 11}
\label{Q11}
\end{figure}

The eleventh question concerns the many worlds interpretation and like the previous question intends to uncover what physicists associate with the interpretation. The many worlds interpretation contains several features, but not all are necessarily known by all physicists. As the previous question, the participants were allowed to pick several options.

Here a clear answer is given, which is that the main association with the many worlds interpretation is the postulate of many worlds. This, of course, is not surprising, since the existence of multiple worlds is expressed in the interpretation's very name. Physicists do not seem familiar with other features of the interpretation, such as locality and the observer being treated as a quantum system. However, almost all of the participants associated no collapse to the theory, which is readily implied by the worlds corresponding to every possible event. 

From the description of the many worlds interpretation, it is worth recalling that what was central to Hugh Everett, who formulated the interpretation, was to solve the measurement problem, and he never used the word "worlds" in his thesis. His focus was on rejecting the collapse postulate.

\begin{figure}[H]
\includegraphics[scale=0.8]{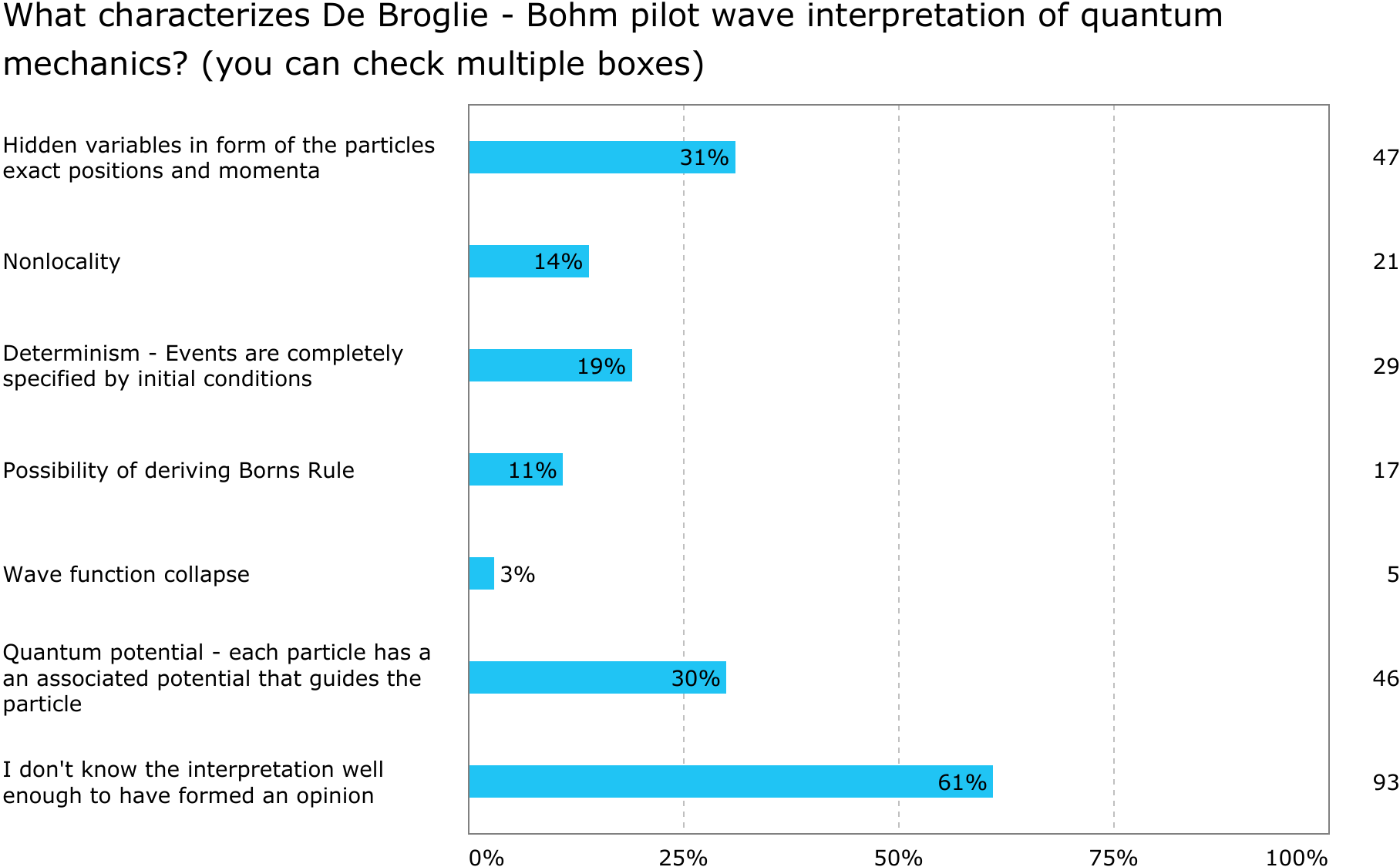}
\caption{Distribution of all the participants answers to question 12}
\label{Q12}
\end{figure}

The twelfth question concerns bohmian mechanics and like the two previous questions intend to uncover the associations made with the interpretation by physicists.

\begin{figure}[H]
\includegraphics[scale=0.8]{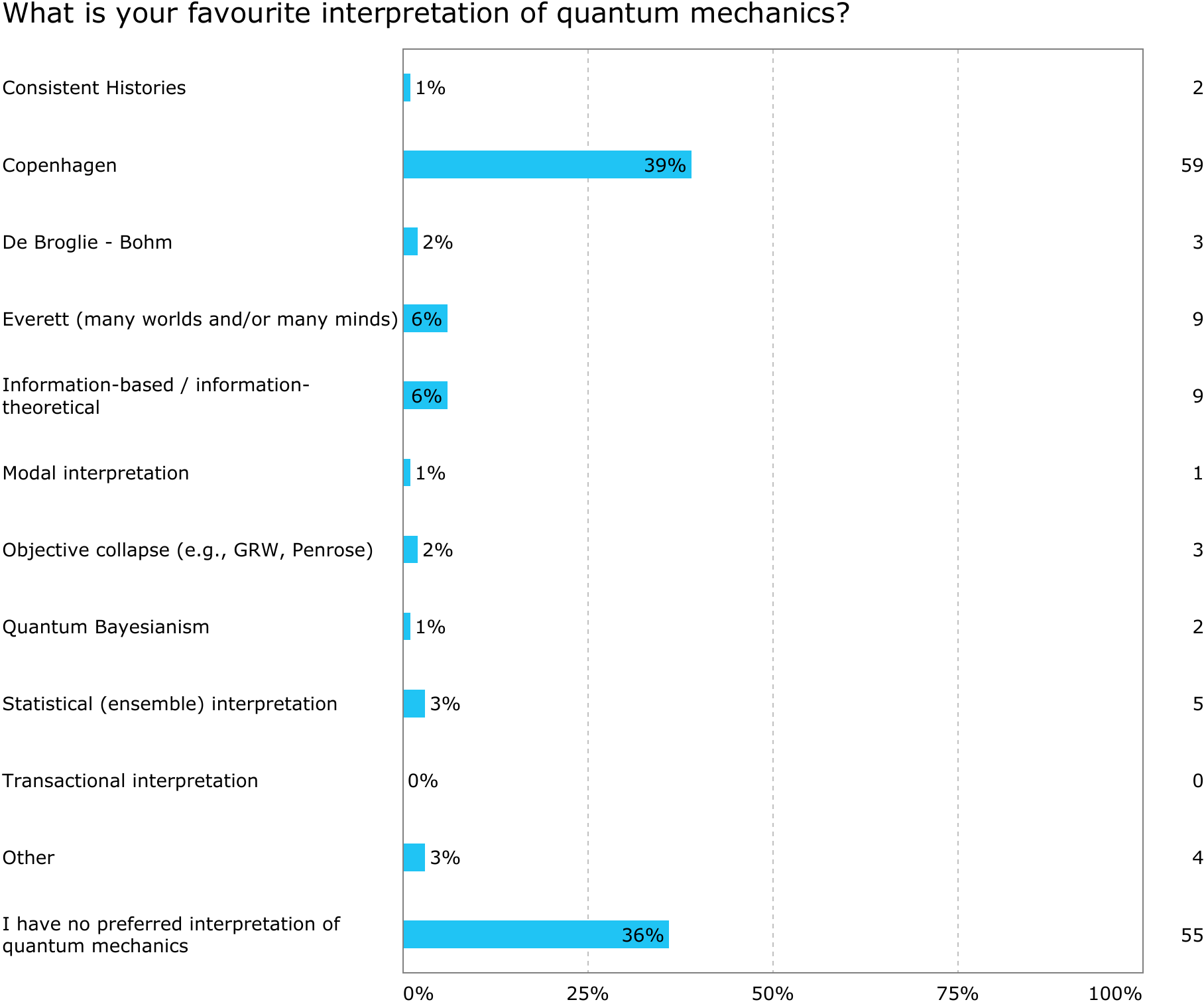}
\caption{Distribution of all the participants answers to question 13}
\label{Q13}
\end{figure}

The thirteenth question can be considered as the main question of the survey since it concerns which interpretation is the most popular today. Besides the several interpretations, a last option of having no preferred interpretation of quantum mechanics is available to incorporate those who do not feel there are any satisfactory interpretations of quantum mechanics, as well as those who have a "shut-up and calculate" approach to quantum mechanics.

The results here show that the Copenhagen interpretation is still by a large margin the preferred interpretation of quantum mechanics with 33 percentage points separating it from the many worlds interpretation and information theoretic approach, which has been said to be an offspring of the Copenhagen interpretation. However, almost as many, 36\%, do not have a preferred interpretation of quantum mechanics. This can be explained by the hypothesis that most physicists are not familiar with, or occupied by quantum interpretation, and either have no preference concerning interpretation or just choose the Copenhagen interpretation by default.

\begin{figure}[H]
\includegraphics[scale=0.8]{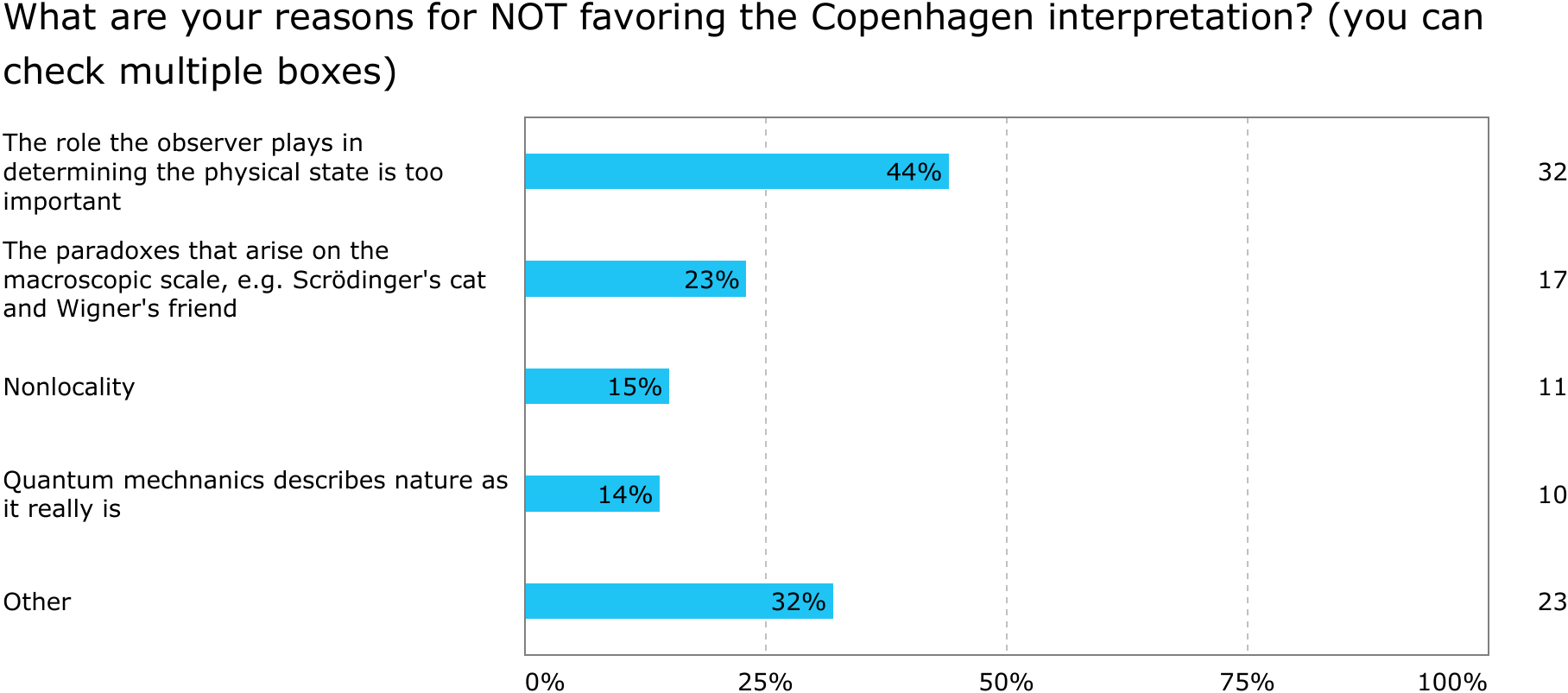}
\caption{Distribution of all the participants answers to question 14}
\label{Q14}
\end{figure}

The fourteenth question concerns the features of the Copenhagen interpretation that seem dissuading. The question was not displayed to every participant, but only those who had not chosen the second option in question 13, i.e. that their preferred interpretation of quantum mechanics is the Copenhagen interpretation, or the seventh option of question 10, i.e. an unfamiliarity with the Copenhagen interpretation. The participants were allowed to pick multiple options to this question.

Of those who do not favor the Copenhagen interpretation, the majority states that it is because of the role the observer plays in the interpretation that they do not favor it. 

A significant part of the participants chose "other", which could imply that a significant reason has been omitted as an option. It is thought that more complex reasons, that are not readily formulated as a survey option, are behind the high frequency of the last option. This is indeed corroborated by some of the comments left by those who chose the last option.

\begin{figure}[H]
\includegraphics[scale=0.8]{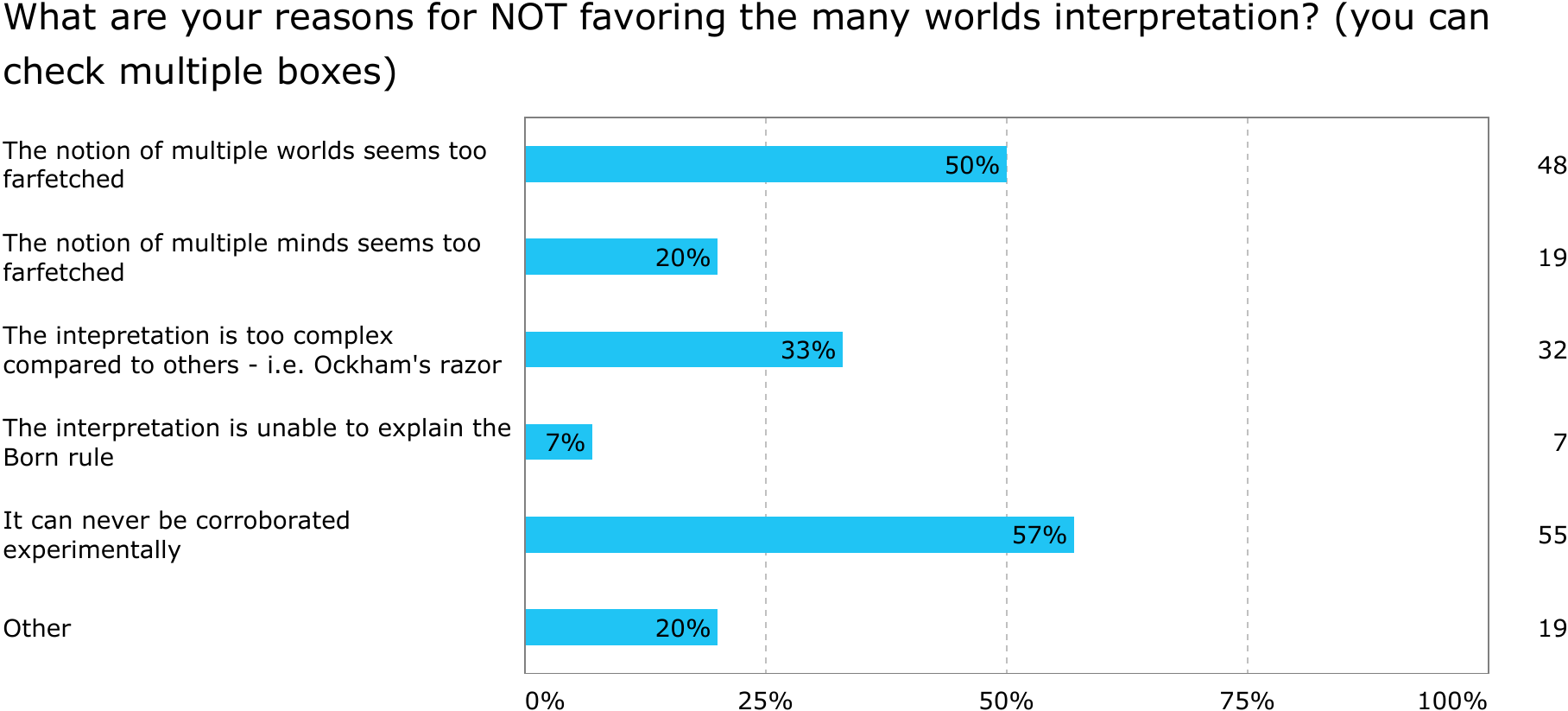}
\caption{Distribution of all the participants answers to question 15}
\label{Q15}
\end{figure}

The fifteenth question concerns which features of the many worlds interpretation seem dissuading. As the previous question, this question was not displayed to the participants who had chosen the fourth option in question 13, i.e. that they favor the many worlds interpretation, or the seventh option of question 11, i.e. an unfamiliarity with the many worlds interpretation. The participants were allowed to pick multiple options to this question.

\begin{figure}[H]
\includegraphics[scale=0.8]{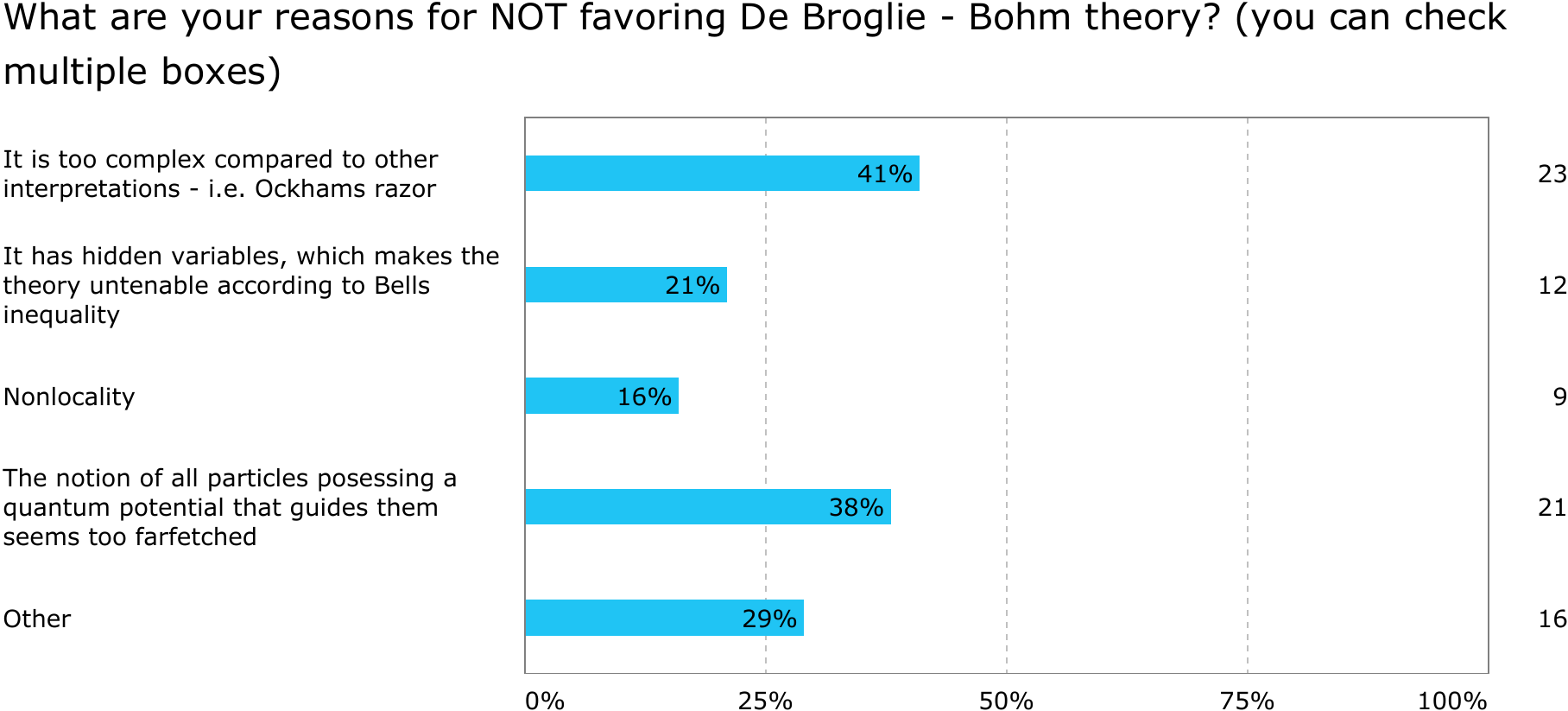}
\caption{Distribution of all the participants answers to question 16}
\label{Q16}
\end{figure}

The sixteenth question concerns what features of bohmian mechanics seem dissuading. As the two previous questions, this question was not displayed to the participants who had chosen the third option of question 13, i.e. that they favor bohmian mechanics, or the seventh option of question 12, i.e. an unfamiliarity with bohmian mechanics. This filter was chosen out of the same reasons for the previous filters. The participants were allowed to pick multiple options to this question.

\begin{figure}[H]
\includegraphics[scale=0.8]{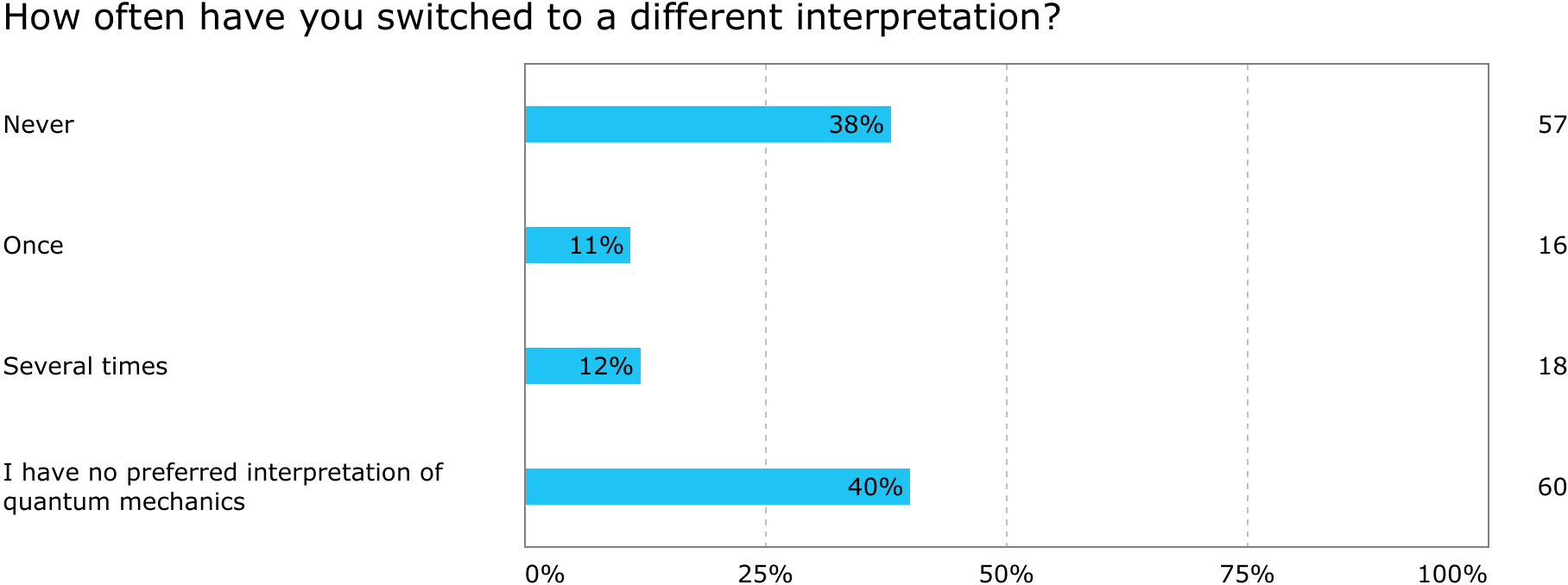}
\caption{Distribution of all the participants answers to question 17}
\label{Q17}
\end{figure}

The seventeenth and last question concern the nature of changing interpretations, whether this is frequently done or never done by physicists. The results show that preferences of interpretations are very inert. Almost 80\% of the participants have never changed interpretation. Another way to regard this results is that the subject of quantum interpretations simply do not occupy the minds of physicists and not given much attention.

\section{Correlations}

To examine various correlations between certain answer options, i.e. if one chose option A in question X there would be a strong likelihood that one would choose option B in question Y. The same scheme was adopted as in Schlosshauer et Al. Various tables can be found in the appendix which illustrates the connections between the different answer options in different ways. Correlations between two answer options A and B were identified by imposing three criteria. 

\begin{itemize}

\item The group of those who chose answer A, who also chose B, must at least contain 21 participants. 

\item The fraction $f$ of those who chose answer A and also answer B to all those who chose A must be higher than a threshold value $T$. 

\item The fraction $f$ must be bigger than the fraction $f'$ of those who chose answer B out of the whole group to the whole group. There must be a gap \textit{G} between these fractions.

\end{itemize} 

The correlations were grouped in two; strong correlations corresponding to $T=80\% \, G=30\% $ and weak correlations corresponding to $T=80 \%  \, G=20 \% $. These values for $T$ and $G$ are different from those used in Schlosshauer et al., to better suit the different sample size. They were also chosen to be "strict" so to exclude seemingly correlations that only stem from pure random choice. In general too much emphasis should not be placed on these correlations, because the validity of them are highly questionable, since they pertain to the complex nature of human opinion, however some of them seem to make good sense, such as a correlation between the favoring the Copenhagen interpretation and never having changed one's preference of interpretation.  

\begin{figure}[H]
\centering
\includegraphics[scale=0.5]{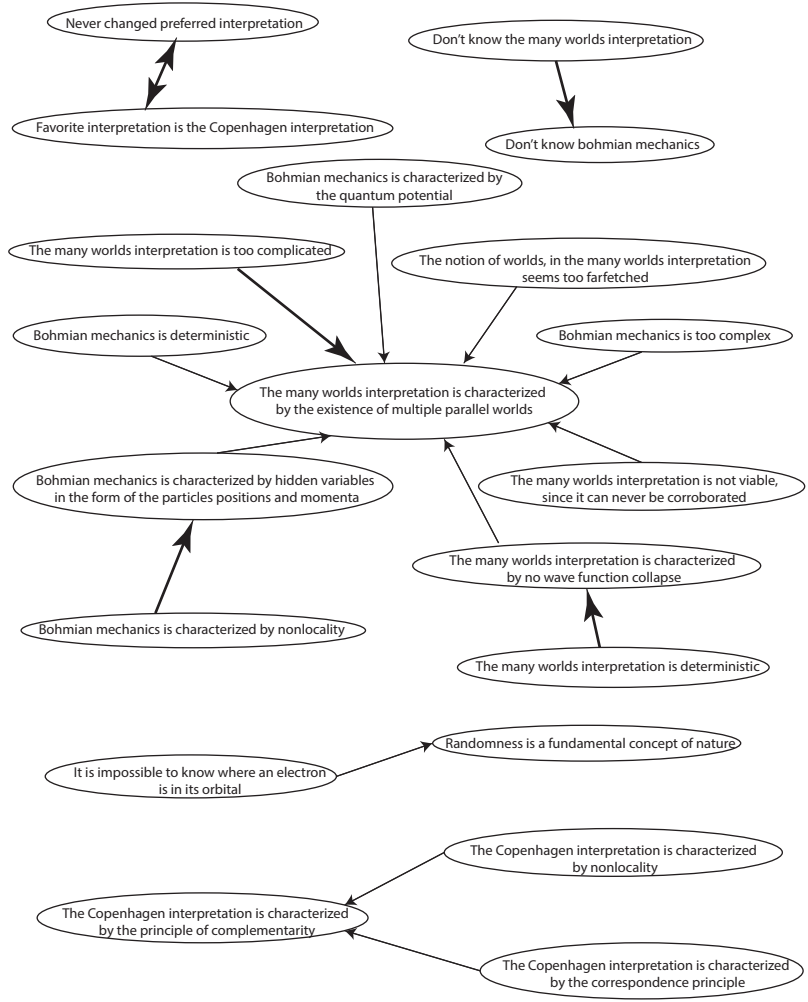}
\caption{The big arrows represent the strong correlations, while the small arrows represent the weak correlation}
\end{figure}

\section{Conclusion}

More and more work is done concerning quantum foundations; investigating basic properties of quantum mechanics, such as Bell’s inequality, or developing new interpretations of quantum mechanics, such as QBism. However, when one regards the results of the survey, it shows that the resurgence the topic has been undergoing in recent times still has not had an impact on the participants being familiar with foundational concepts. This is seen from the answers to the questions concerning Bell’s inequality and the measurement problem, where a minority of the participants had a proper grasp of these topics. The results of the survey, in contrast, also reveal that most of the participant feel that the question of interpretation is an important one, seen from the answers to the questions of whether quantum mechanics needs an interpretation. This seems as quite a validation of the whole research area concerning quantum foundations from the general physics community.

Of course one should be very cautious in extrapolating the answers from the participants of the survey, to represent the whole of the physics community. Even though the sample size in this survey is significantly larger than other surveys conducted in relation to the same topic, the sample size is still too small\footnote{This assertion is made solely on statistics, and the goal of having a 95\% confidence interval. The proper sample size, of course, depends on the population size, which is not readily estimated, but when the population is everyone with a master's degree in physics, 149 cannot give the desired confidence interval.}. Furthermore, the questions and the answer options in a survey, cannot capture various nuances and it is easier to hide one's ignorance in relation to various issues when the survey is in a multiple-choice format.

A better way to survey the attitudes of the physics community concerning foundational issues of quantum mechanics would be to have the participants describe various concepts. Such as having the participants describe their understanding of the measurement problem or the Copenhagen interpretation. The Copenhagen interpretation would be of particular interest, since today it is recognized by several historians that the Copenhagen interpretation is not homogeneous view of quantum mechanics, and there are indeed several Copenhagen interpretations. A survey highlighting this inconsistency in the physics community would be a remarkable achievement. The answers to such surveys are much more difficult to analyze, than those of multiple-choice format, especially if the sample size is large. Furthermore surveys of such nature would probably deter a lot of would-be participants from answering the survey since it would be more laborious, therefore limiting the sample size. Nonetheless, such a survey would be able to reveal much and substantially corroborate many of the conclusions drawn from this survey.  

\pagebreak

\section*{Acknowledgements}

First of all a big "thank you" is in place to my supervisor Kristian Hvidtfelt Nielsen. I know I am not the easiest person to work with, because of my erratic work method and lack of organization, but you have had the right sense of when to push and when to give me space. This was, and is, very much appreciated and I hope that is not lost on you.

I would also like to thank Klaus Mølmer for the guidance and support he has provided, whether it be suggesting my supervisor or in getting me more participants for my survey. Talking to you concerning quantum foundations has been one of the most enjoyable parts of this process, this was especially highlighted by your enthusiasm, which is admirable and contagious. It is best described by the first time I came to you to talk about interpretations. I clearly recall uttering the words "interpretations of quantum mechanics" and you went off talking about the subject for half an hour, finishing with "I don't know if that answered your question" even though I never got to ask my question.

A thank you should also be given to Magnus Johan Aarslev, Daniel Østergaard Andreasen and Mehmet Serdar Yilmaz for helping me test the survey, Brian Julsgaard for helping me clarify some features of quantum mechanics and of course to all those who participated in the survey. Without your participation, none of the following would have been of much use.

\thispagestyle{empty}

\appendix

\section{Appendix}

\begin{figure}[H]
\includegraphics[scale=0.70]{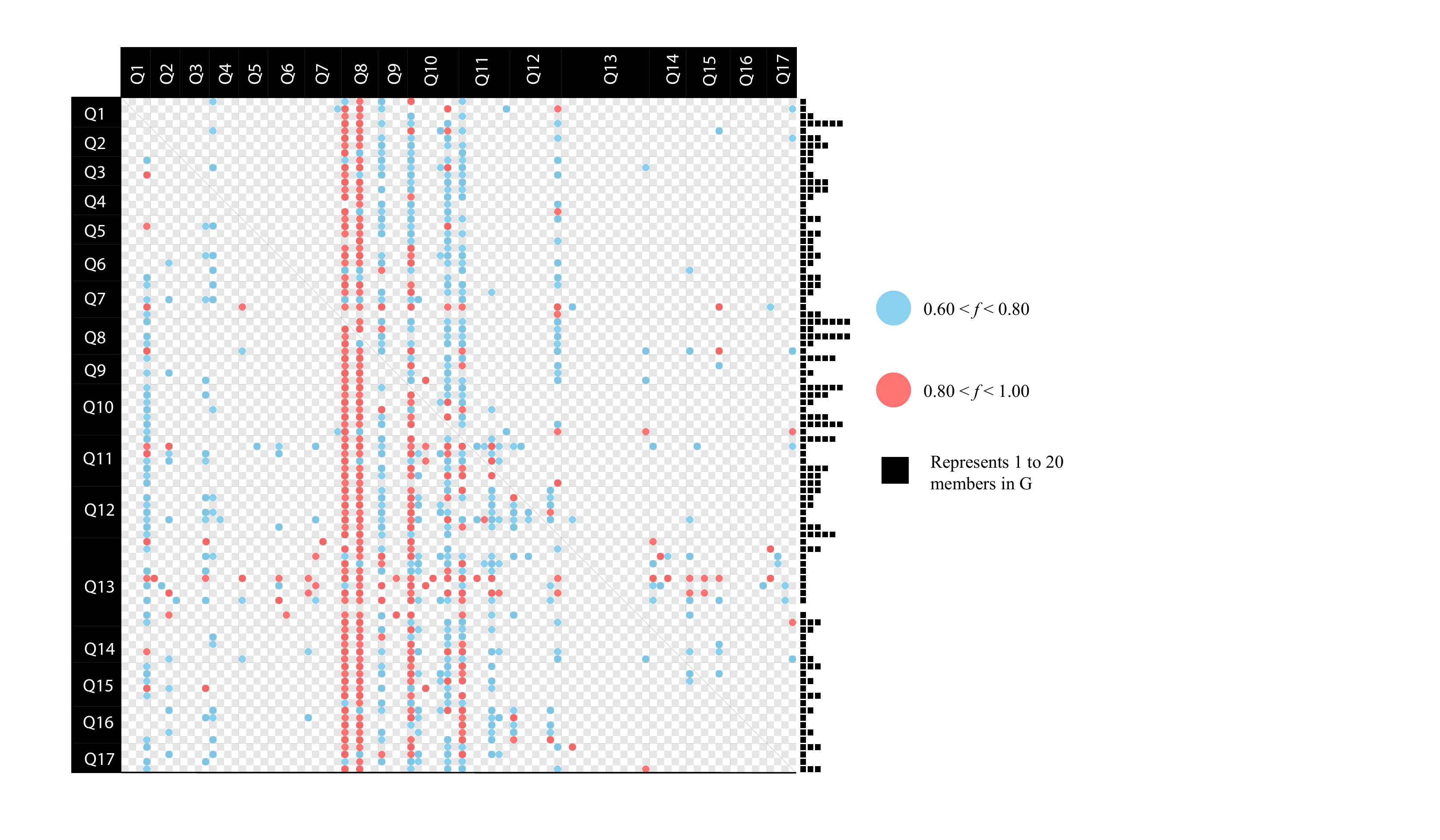}
\caption{The table shows the value of the fraction $f$ between various answer options. Each row corresponds to an answer option and shows what those who have chosen this option have answered in other questions along the row. The squares on the side indicate how many of the participants chose the option corresponding to the particular row. One square means between 1-20, two squares between 21-40 and so forth. Zero squares mean no one picked the option. The diagonal is crossed, since it carries no information.}
\end{figure}

\pagebreak

\begin{figure}[H]
\includegraphics[scale=0.70]{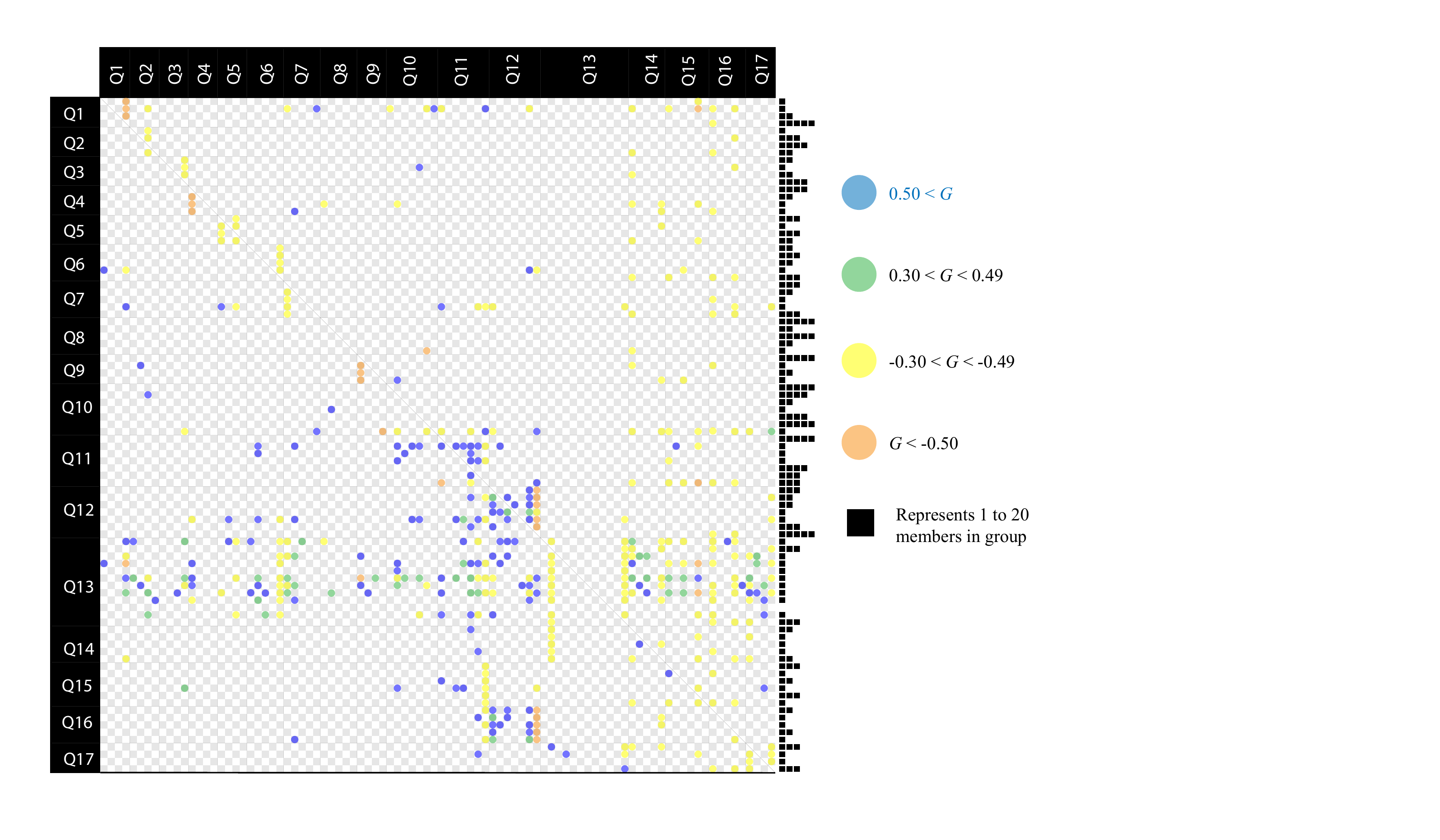}
\caption{The table shows the value of the gap $G$ for various answer options.}
\end{figure}

\end{document}